# The effect of nudging personal and injunctive norms on the trade-off between objective equality and efficiency

Steven J. Human and Valerio Capraro

Middlesex University London

**Abstract**

We report three pre-registered studies (total N=1,799) exploring the effect of nudging personal and injunctive norms in decisions that involve a trade-off between objective equality and efficiency. The first two studies provide evidence that: (i) nudging the personal norm has a similar effect to nudging the injunctive norm; (ii) when both norms are nudged towards the same direction, there is no additive effect; (iii) when the personal norm and the injunctive norm are nudged towards opposite directions, some people tend to follow the personal norm, while others tend to follow the injunctive norm. Study 3 tests whether these two classes of people, those who tend to follow the injunctive norm and those who tend to follow the personal norm, map onto the two sub-dimensions of Aquino and Reed's moral identity scale. We find partial evidence of this hypothesis: people higher in the symbolisation dimension are weakly more likely to follow the injunctive norm; however, we do not find any evidence that people higher in the internalisation dimension are more likely to follow the personal norm.

**Introduction**

*Homo Economicus* has been the centre of debate within the social sciences for centuries. Defined as a human who "inevitably does that by which he may obtain the greatest amount of necessaries, conveniences, and luxuries, with the smallest quantity of labour and physical self-denial with which they can be obtained" (Mill, 1836), *Homo Economicus* has formed the basis of neoclassical economic theory.

However, in contrast to the theoretical predictions of this model, behavioural experiments have proved that, in several situations, people's preferences go beyond their economic interests alone, even in one-shot, anonymous, economic games, where there are no direct or indirect incentives to deviate from money-maximisation. For instance, a significant number of dictators give some money in the dictator game[1]. Another observed regularity is that in the ultimatum game,[2] a significant number of responders reject offers greater than zero and a significant number of proposers offer half of the endowment (rather than the smallest available increment above zero) (Camerer, 2011). Another game in which people's behaviour is inconsistent with money-maximisation is the one-shot prisoner's dilemma[3]. Empirical evidence indicates that a significant proportion of individuals choose to cooperate in this setting, even though it is not individually optimal (Rapoport and Chammah, 1965; Ledyard, 1995).

An influential stream of research, which started in the late 1990s, has seen these empirical regularities explained using outcome-based (i.e. explicitly linked to monetary payoffs) social preferences. These models argue that participants are impacted negatively in a psychological sense, meaning that they incur costs (disutility), if payoffs are unequal ("inequity aversion") (Bolton & Ockenfels, 2000; Fehr & Schmidt, 1999) or are affected positively, or gain utility, when the total payoff of all players involved in the interaction increases (this could be altruistic or driven by preferences for socially efficient outcomes) (Charness & Rabin, 2002; Engelmann & Strobel, 2004). These models are well regarded and prevalent in a plethora of literature, but they are only effective in explaining prosocial behaviour in terms of outcome-based preferences. In fact, a number of studies in the past two decades have shown that individuals do not always act in a way that is consistent with outcome-based utility functions.

For example, a substantial proportion of people act honestly even when it lowers their payoff, but will instead choose to maximise their payoff when lying is not necessary to this end (Gneezy, 2005). This presents clear and direct evidence that there is some intrinsic cost to lying that is not explained by outcome-based preferences. More evidence of this has been presented by Biziou-van-Pol, Haenen, Novaro, Liberman and Capraro (2015), who showed that cooperation in the one-shot prisoner's dilemma and altruistic giving in the dictator game are both positively correlated to truth-telling in a "Pareto white lie" situation, whereby lying would increase the payoff of all participants involved. These results provide evidence for some common underlying non-outcome-based motivation that drives honesty, cooperation, and altruistic behaviour, motivation that the authors call "preference for doing the right thing". In line with this view, recent work has shown that instructions using morally loaded language can significantly influence behaviour in several games, including the ultimatum game, the dictator game, and the trade-off game[4] (Eriksson et al. 2017; Capraro & Rand, 2018; Tappin & Capraro, 2018; Capraro & Vanzo, 2019; Capraro et al. 2019; Capraro, Rodriguez-Lara, & Ruiz-Martos, 2020; Capraro, Jordan, & Tappin, 2020), by changing people's perception about what is the right thing to do in the given situation, which is also known as "personal norm" (Schwartz, 1994). In sum, this stream of literature shows that in several one-shot, anonymous, economic games, the behaviour of some people goes beyond outcome-based preferences, but can be rationalised by a "generalised morality preference" for following their personal norm.

---
[1] The dictator game involves two players, one is the dictator and the other is the recipient. The dictator divides a set amount between himself and the recipient. The recipient has no choice and only gets the amount that the dictator decides to give.
[2] The ultimatum game is a sequential two-player game with the first player being the proposer and the second player the respondent. The first player is tasked with splitting a sum of money between both players. The second player chooses to accept or reject the offer. If the second player rejects the offer both players receive nothing.
[3] The prisoner's dilemma is a simultaneous two-player game in which players choose either to cooperate or defect. If both cooperate the overall payoff is maximised; if both defect the overall payoff is minimal; if one player cooperates while the other defects, the cooperator receives the worst possible payoff while the defector receives the highest possible payoff.

Alongside the personal norm, other works have shown the importance of the "injunctive norm". The injunctive norm is defined as what one ought to do and is based on one's belief of what *others* think the right thing for them to do is (Cialdini, Reno, & Kallgren, 1990). Several works have provided evidence that, in some contexts, people seem to follow what they believe to be the injunctive norm in those contexts (Assar, Nyberg, & Weibull 1999; Bendor & Swistak, 2001; Bicchieri, 2005; Camerer & Fehr 2004; Cialdini, Reno, & Kallgren, 1990; Coleman, 1994; Hechter & Opp, 2001; Kandori 1992).

However, in the specific setting of one-shot, anonymous, economic games, results are mixed and whether the injunctive norm even plays a role in these contexts is still debated. Using variations of the dictator game, Krupka and Weber (2013) presented correlational evidence that people seem to follow what they believe to be the injunctive norm. However, Schram and Charness (2015) failed to find a causal link between the injunctive norm and dictator game choices in anonymous games: they found that informing dictators about what other dictators think to be the right thing affects dictators' choices only when these choices are made public. Dimant, Van Kleef and Shalvi (2020) found that norm-based messages giving cues about the injunctive norm had no effect on honesty in a mind game, because, in their case, they failed to manipulate the norm. On the other hand, Bicchieri and Chavez (2010) manipulated the injunctive norm in variants of the ultimatum game and found that this does have an effect on proposers' offers. In sum, previous work provides mixed evidence about the causal role of the injunctive norm in one-shot, anonymous, economic games.

In this paper, we contribute to this literature along two dimensions. First of all, we study the causal role of the injunctive norm in the trade-off game. We chose this game for three reasons. The first is that the trade-off game formalises situations where there is a conflict between objective equality and efficiency and therefore it is a good representation of many situations faced by people who are responsible for the distribution of resources, making the game important from a practical perspective. The second is that, as reviewed earlier, previous work has not explored the effect of nudging the injunctive norm in the trade-off game; therefore, our work presents a novel contribution. The third is that previous work has shown that nudging the personal norm in the trade-off game does have an effect (Capraro & Rand, 2018; Tappin & Capraro, 2018; Capraro, Rodriguez-Lara, Ruiz-Martos, 2020; Capraro, 2020); therefore, we can also compare the effect of nudging the injunctive norm with the effect of nudging the personal norm. This leads us to the second novel contribution of this work: the comparison between the effect of nudging the personal norm and the effect of nudging the injunctive norm. Indeed, to the best of our knowledge, the injunctive norm and the personal norm have not been compared in the literature, neither in general, nor in the trade-off game in particular.[5]

---

[4] The trade-off game is a three-player game. Player A decides between two payoff distributions, one that is equal, for example (13,13,13), and one that is efficient, for example (13,23,13). Player B and C make no decisions and take no actions. The first component corresponds to the payoff of Player A, the second component corresponds to the payoff of Player B, the third component corresponds to the payoff of Player C.

We report three pre-registered studies (total N = 1,799). The first two studies provide evidence that nudging the injunctive norm in the trade-off game does have an effect, and this effect is, in size, similar to the effect of nudging the personal norm. Moreover, when the two norms point towards different options, some participants prefer to follow the injunctive norm while others tend to follow the personal norm. The latter result suggests that there might be some underlying individual characteristics driving people towards the injunctive or the personal norm. Study 3 aims to investigate two such candidates, respectively, the symbolisation and the internalisation dimensions of the moral identity scale (Aquino & Reed, 2002). As we predicted, this study shows that people high in the symbolisation dimension of the moral identity scale – who like being seen as moral – are more likely to follow the injunctive norm. However, contrary to our predictions, we do not find any clear relation between people who are high in the internalisation dimension of the moral identity scale – who like to see themselves as moral – and their tendency to follow the personal norm.

## Study 1

The design, the hypotheses, the exclusion criteria, the sample size, and the analyses were preregistered at: https://aspredicted.org/blind.php?x=5rk5yn.

**Methods**

The experiment was conducted between the 5th and the 6th of May 2020 on Amazon Mechanical Turk (AMT; Paolacci, Chandler, & Ipeirotis, 2010), through a survey built in Qualtrics. Participation in the study was restricted to AMT workers based in the US with at least 90% acceptance rate in previous HITs. Once the survey was completed, we excluded duplicate Turk IDs and duplicate IP addresses before analysis began. In the cases where we found duplicates, we only kept the first observation (determined by the starting date) and we discarded the remaining cases. These accounted for less than 3% of the total observations.

Participants were randomly assigned to one of five conditions. Each condition was a slight variation of the same trade-off game (Capraro & Rand, 2018; see Footnote 4 for the definition). In the *baseline* condition ("BL", N=175), the allocation (13,13,13) was termed "Option A", whereas the allocation (13,23,13) was termed "Option B". In the *personal norm equal* condition ("PN_equal", N=162), the allocation (13,13,13) was termed "Fair", whereas the allocation (13,23,13) was termed "Unfair". We name this condition "*personal norm equal*" because previous work has shown that the vast majority of people believe that choosing to be fair is the morally right thing to do in this case (Capraro & Rand, 2018). The *personal norm equal + injunctive norm equal* condition ("PN_equal + IN_equal", N=158) is identical to the *personal norm equal* condition, but, right before making their choice, participants are told: "we asked 10 participants who played before you what they think the morally right thing to do is. Nine of these ten participants (90%) declared that choosing for all players to get the same payoffs is the morally right thing to do". The *personal norm equal + injunctive norm efficient* condition ("PN_equal + IN_efficient", N=166), is similar to the previous condition, but this time participants are told: "Nine of these ten participants (90%) declared that choosing to maximize the sum of the payoffs of all players is the morally right thing to do". Note that this methodology does not involve deception: Capraro and Rand (2018) collected data about the moral judgments in the *PN_equal* condition. Finally, the *injunctive norm efficient* condition ("IN_efficient", N=171) was identical to the PN_equal + IN_efficient condition, but the options were referred to as "Option A" and "Option B". We note that there is a discrepancy with the pre-registration: because of an internal mistake, in the pre-registration we did not pre-register the *IN_efficient* condition, but we pre-registered the *IN_equal* condition, where the injunctive norm was placed on the equal choice.

---

[5] We are aware of a handful of studies comparing other types of norms. The descriptive norm and the injunctive norm have been compared in the well-known field experiment involving litter in a street carried out by Cialdini, Reno and Kallgren (1990). From this, they deduced that when the injunctive norm (I ought not litter) is in conflict with the descriptive norm (people are clearly littering) individuals follow the descriptive norm. Bicchieri and Xiao (2009) conceptually replicated this finding in a lab experiment. Capraro and Rand (2018) instead compared the descriptive norm and the personal norm in the trade-off game. They did so by having subjects play the game with morally loaded language, so that to change people's perception of what is the right thing to do. They found that giving incongruent information about the behaviour of other people (the descriptive norm) has very little effect, as people tend to keep following their personal norm.

We included comprehension questions before the decision to ascertain that participants understood the game and the implications of their choices. Those who did not respond correctly to all questions were excluded. After making their decision, participants were asked a set of demographic questions, at the end of which they received the completion code needed to claim the payment. After collecting the data, we computed the bonuses and paid them, on top of the participation fee (50c for 4-minute survey). Each participant played as Player A; Player B and Player C were randomly selected by a pool of participants who participated in other experiments, and were paid according to Player A's decision. This study does not contain deception. We refer to the Appendix for full experimental instructions.

**Hypotheses**

As mentioned in the Introduction, previous work found that nudging the injunctive norm has an effect in the ultimatum game (Bicchieri & Xiao, 2009), but not in the dictator game (Schram & Charness, 2015). Since the trade-off game is closer to the dictator game than it is to the ultimatum game (e.g., the trade-off game is a unilateral game, where only one player makes a decision and the other players are passive), we hypothesized that in the trade-off game the injunctive norm would play very little role. From a theoretical perspective, our leading idea was that the injunctive norm starts playing a role only when there is more than one decision-maker (as in the ultimatum game), because in this case decision makers need to take into account what others approve or disapprove of. To reflect this intuition, we pre-registered two hypotheses. The first of these two hypotheses was divided in two sub-hypotheses, one of which concerned the treatment *IN_equal*. Since, as mentioned above, in the actual experiment we did not conduct the *IN_equal* condition, we focus on the second part of the first hypothesis and on the second hypothesis:

1.1 People are more likely to choose the equal option in the *PN_equal* condition compared to the *Baseline* condition.
1.2 People are equally likely to choose the equal option in the *PN_equal* condition, compared to both the *PN_equal + IN_efficient* condition and the *PN_equal + IN_equal* condition.

Note that Hypothesis 1.1 does not involve the injunctive norm and, indeed, is not new (Capraro & Rand, 2018; Tappin & Capraro, 2018; Capraro, Rodriguez-Lara, & Ruiz-Martos, 2020; Capraro, Jordan, & Tappin, 2020; Capraro, 2020).

**Results**

Figure 1 reports the average percentage of participants who choose the equal option. For completeness, we report also the results of the non-preregistered condition, *IN_efficient*. Error bars represent 95% confidence intervals. Horizontal bars reports coefficients and p-values of pairwise logistic regression (our pre-registered statistical test). Table 1 summarises regression results.

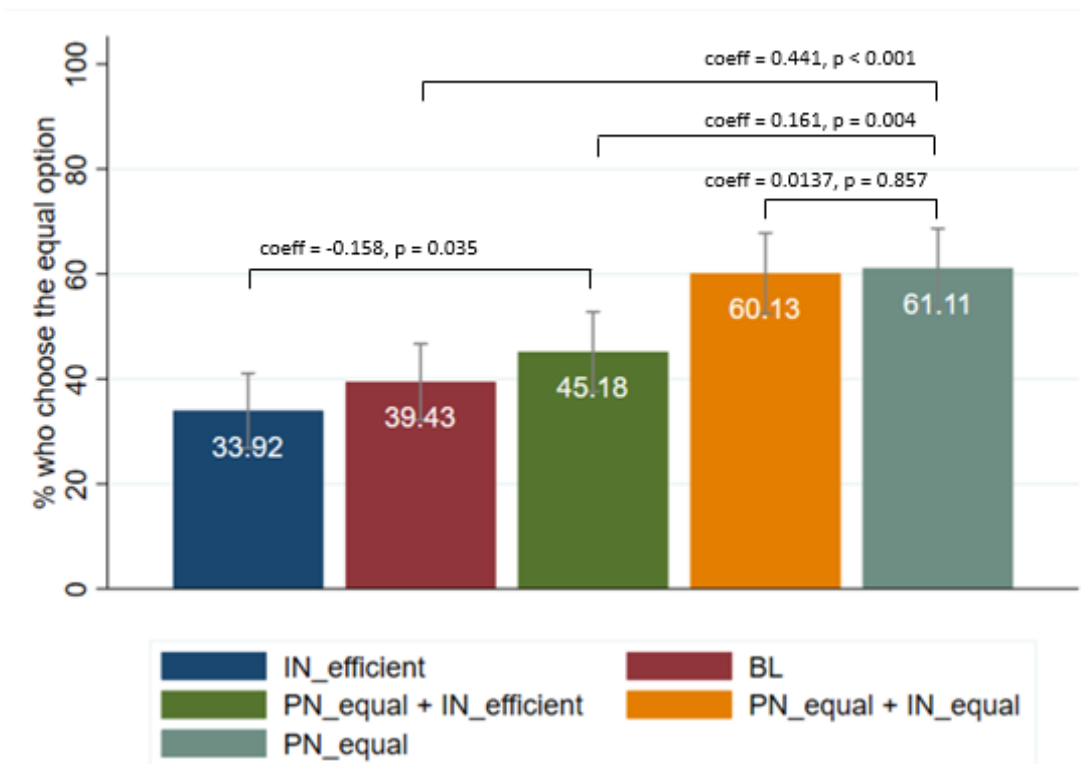

*Figure 1. Percentage of participants who choose the equal option in each condition. Error bars represent 95% CI and statistics refer to results of logistic regressions.*

### Test of hypothesis 1.1

In the *Baseline* we find that 39.4% of the participants choose the equal choice. In the *PN_equal* condition, 61.1% of the participants choose the equal choice. Therefore, in line with hypothesis 1.1 and in line with previous work, we find that making the personal norm salient highly impacts people's behaviour (coeff = 0.441, p < .001).

### Test of hypothesis 1.2

We now compare the proportion of equal choices in the *PN_equal + IN_equal* condition with that in the *PN_equal* condition. We find that these are not statistically different (coeff = 0.0137, p = 0.857). This is in line with hypothesis 1.2, and suggests that promoting the injunctive norm on top of the personal norm does not have any sizeable additional effect. On the other hand, if we compare the *PN_equal + IN_efficient* condition with the *PN_equal* condition, we do find a statistically significant difference (coeff = 0.161, p = 0.004). We did not predict this result, which suggests that when the injunctive norm conflicts with the personal norm, some, and possibly all, participants follow the injunctive norm.

*Non-preregistered exploratory analysis*

To understand whether all participants follow the injunctive norm or only some of them, we compare the *IN_equal* condition with the *PN_equal + IN_efficient* condition. In doing so, we find that they are statistically different (coeff = -0.158, p = 0.035). This suggests that when the injunctive norm is misaligned with the personal norm, some participants follow the injunctive norm while others follow the personal norm.

|  | **PN_equal vs BL** | **PN_equal vs PN_equal + IN_efficient** | **PN_equal vs PN_equal + IN_equal** | **IN_equal vs PN_equal + IN_efficient** |
|---|---|---|---|---|
| **Treatment** | 0.441*** | 0.161** | 0.0137 | -0.158* |
|  | (0.112) | (0.0561) | (0.0763) | (0.0748) |
| **Constant** | -1.751*** | -0.355 | 0.383 | -0.0355 |
|  | (0.456) | (0.199) | (0.291) | (0.215) |
| **Observations** | 337 | 328 | 320 | 337 |

*p-Value<0.1, **p-value<.0.05, and ***p-value<.0.01.
(Standard errors in parentheses)

*Table 1. Pairwise logistic regressions. Treatment is a dummy that takes value of 1 for the first treatment in the comparison and 0 otherwise.*

**Discussion**

Apart from replicating previous work regarding the effect of nudging the personal norm on the trade-off game, Study 1 reports a curious asymmetry such that, when both the injunctive norm and the personal norm nudge towards the same option (e.g., they both promote the equal choice), then their effect is not additive; however, when they nudge towards different options, some people tend to follow the injunctive norm and some people tend to follow the personal norm. Therefore, contrary to our initial predictions, nudging the injunctive norm seems to have an effect on the trade-off game. Study 2 aims to directly test this hypothesis. Additionally, Study 2 aims at comparing the effect of nudging the injunctive norm with the effect of nudging the personal norm.

## Study 2

The design, the hypotheses, the exclusion criteria, the sample size, and the analyses were preregistered at: https://aspredicted.org/blind.php?x=zf57xx.

**Methods**

The experiment was conducted between the 28th and the 29th of May 2020 on AMT. As preregistered, participation in the study was restricted to AMT workers based in the US with at least 90% acceptance rate in previous HITs. Once the survey was completed, we excluded duplicate Turk IDs and duplicate IP addresses before analysis began. In the cases where we found duplicates, we only kept the first observation (determined by the starting date) and we discarded the remaining cases. These accounted for less than 4% of the total observations. None of the participants in Study 1 were allowed to participated in this study.

As preregistered, participants were randomly assigned to one of three conditions. Each condition is a slight variation of the same trade-off game. In the *baseline* condition ("BL", N=187), the allocation (5,5,5) was termed "Option A", whereas the allocation (5,10,5) was termed "Option B". In the *personal norm equal* condition ("*PN_equal*", N=172), the allocation (5,5,5) was termed "Fair", whereas the allocation (5,10,5) was termed "Unfair". The *injunctive norm equal* condition ("*IN_equal*", N=180) is identical to the baseline condition, but, right before making their choice, participants are told: "we asked 10 participants who played before you what they think the morally right thing to do is. Nine of these ten participants (90%) declared that choosing for all players to get the same payoffs is the morally right thing to do". Note that this methodology does not involve deception: Capraro and Rand (2018) collected data about the moral judgments in the *PN_equal* condition.

After making their decision, participants were asked a set of demographic questions. The payment procedure was the same as Study 1. This study does not use deception. We refer to the Appendix for full experimental instructions.

**Hypotheses:**

In light of the results of Study 1, we pre-registered the following hypotheses:

2.1 Nudging either norm increases the proportion of people choosing the equal choice, compared to the baseline.
2.2 The difference between nudging the personal norm and nudging the injunctive norm is negligible.

**Results**

Figure 2 reports the average percentage of participants who choose the equal option. Error bars represent 95% confidence intervals. Horizontal bars report coefficients and p-values of pairwise logistic regression (our pre-registered test). Table 2 summarises regression details.

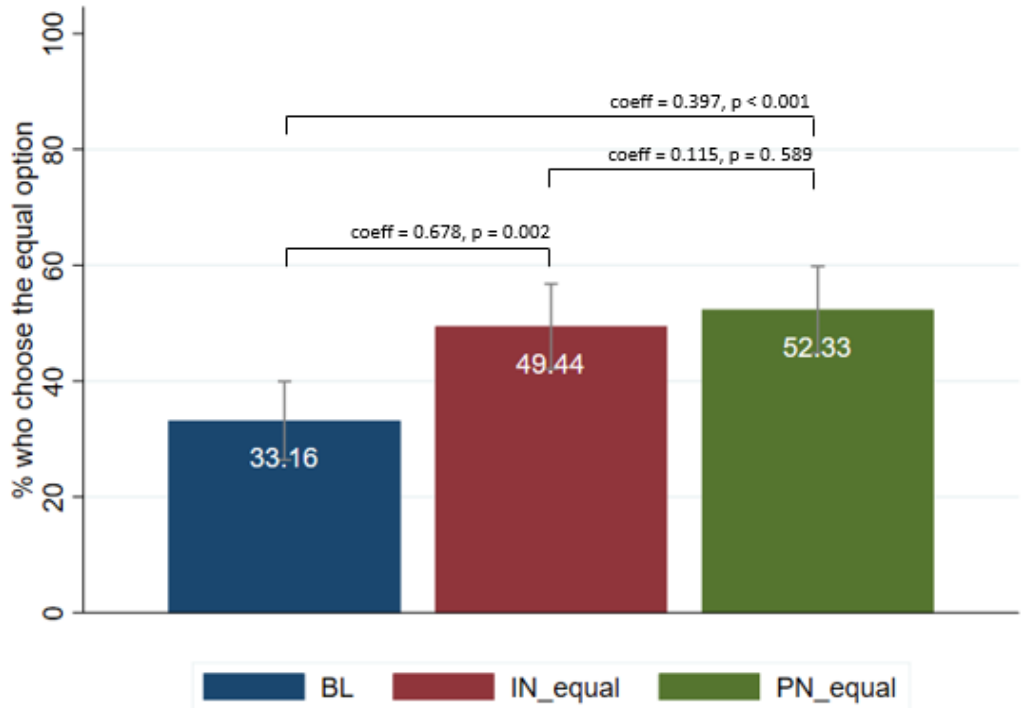

*Figure 2. Percentage of participants who choose the equal option in each condition. Error bars represent 95% CI and statistics refer to results of logistic regressions.*

### Test of hypothesis 2.1

In the *baseline* condition we find that 33.2% of the participants choose the equal choice. In the *PN_equal* condition, 52.2% of the participants choose the equal choice. Therefore, in line with hypothesis 2.1 and in line with previous work, we found that making the personal norm salient highly impacts people's behaviour (coeff = 0.397, p < .001). In the *IN_equal* condition, we found that 49.4% of participants chose the equal choice and, therefore, in line with hypothesis 2.1, nudging the injunctive norm also significantly changes people's behaviour in the predicted direction (coeff = 0.678, p = 0.002).

### Test of hypothesis 2.2

We now compare the proportion of equal choices in *IN_equal* condition with that in the *PN_equal* condition. In doing so, we found that these are not statistically different (coeff = 0.115, p = 0.589). This shows that, in line with hypothesis 2.2, nudging the injunctive norm or nudging the personal norm in the trade-off game have similar effects.

|  | PN_*equal* vs IN_*equal* | PN_*equal* vs Baseline | IN_*equal* vs Baseline |
| --- | --- | --- | --- |
| **Treatment** | 0.115 | 0.397*** | 0.679** |
|  | (0.213) | (0.109) | (0.215) |
| **Constant** | -0.253 | -1.098*** | -1.380*** |
|  | (0.542) | (0.245) | (0.345) |
| **Observations** | 352 | 359 | 367 |

*p-Value<0.1, **p-value<.0.05, and ***p-value<.0.01.
(Standard errors in parentheses)

*Table 2. Pairwise logistic regressions. Treatment is a dummy that takes value of 1 for the first treatment in the comparison and 0 otherwise.*

**Discussion**

In line with our pre-registered hypotheses, Study 2 shows that nudging the personal norm and nudging the injunctive norm in the trade-off game has a similar effect. One possible explanation for these results and those of Study 1 is that people have general moral preferences for following either the personal norm or the injunctive norm. When one of these norms is nudged, then all people with moral preferences move towards this norm. Therefore, when both norms are nudged, but they target the same choice, nothing else happens and there is no additive effect. However, when the norms are misaligned, people turn to norm-specific preferences: some prefer the injunctive norm and others prefer the personal norm. Study 3 sets out to investigate the personality characteristics of those who prefer the injunctive norm and those who prefer the personal norm. We use Aquino and Reed's (2002) morality scale to try and understand if there is a link between the underlying moral identity of participants and their preference for following either the injunctive or personal norm. We choose this scale because it contains two subscales which measure "symbolization" and "internalisation". Internalisation refers to the degree to which moral traits are central to one's self-concept, whereas symbolisation refers to the degree to which these moral traits are reflected in public choices and/or actions in identifiable social settings. These two subscales intuitively correlate with the tendency to follow the personal and injunctive norms, respectively.

<div align="center">

**Study 3**

</div>

The design, the hypotheses, the exclusion criteria, the sample size, and the analyses were preregistered at: https://aspredicted.org/blind.php?x=it59s3.

**Methods**

The experiment was conducted between the 24th and the 26th of June 2020 on AMT. As in the previous studies, participation in the study was restricted to AMT workers based in the US with at least 90% acceptance rate in previous HITs. Once the survey was completed, we excluded duplicate Turk Ids and duplicate IP addresses before analysis began. In the cases where we found duplicates, we only kept the first observation (determined by the starting date) and we discarded the remaining cases. These accounted for less than 10% of the total observations. None of those who participated in the previous studies were allowed to participate in this study.

Participants were randomly divided between two conditions: the *Baseline* condition ("BL", N=226), and the *personal norm equal + injunctive norm efficient* condition ("PN_equal + IN_efficient", N=202), as in the previous studies. We used comprehension questions before the decision to confirm that participants understood the game and the implications of the available choices. Those who failed to correctly answer all comprehension questions were excluded from the survey. After making their decision, participants were asked a set of demographic questions, after which they took the 10-item moral identity scale (Aquino & Reed, 2002). These responses were used to construct internalisation and symbolisation scores. The payment procedure was the same as the previous studies. This study does not use deception. We refer to the Appendix for full experimental instructions.

**Hypotheses**

We pre-registered two hypotheses, reflecting our intuition that people who are high in the symbolisation dimension are more likely to follow the injunctive norm, whereas people who are high in the internalisation dimension are more likely to follow the personal norm.

3.1 The interaction between symbolisation score and trade-off game condition significantly predicts whether participants will choose the efficient choice, such that, in the *PN_equal + IN_efficient* condition, participants that are high in the symbolisation dimension will be more likely to follow the injunctive norm and choose the efficient choice.
3.2 The interaction between internalisation score and trade-off game condition significantly predicts whether participants will choose the equal choice, such that, in the *PN_equal + IN_efficient* condition, participants that are high in the internalisation dimension will be more likely to follow the personal norm and choose the equal choice.

**Results**

*Test of hypothesis 3.1*

As pre-registered, we use logistic regression to estimate treatment effect (a binary variable with *BL* = 0 and *PN_Equal + IN_efficient* = 1), the effect of symbolisation score (calculated by summing up all the scores in the internalisation questions in the moral identity scale) as well as the interaction of the symbolisation score and the treatment variable. Hypothesis 3.1 corresponds to a significant negative interaction. Table 3, first column, provides weak support for this hypothesis, as the interaction is negative and weakly significant (coeff = -0.163, p = 0.089).

*Test of hypothesis 3.2*

We use logistic regression to estimate treatment effect, the effect of internalisation score (calculated by summing up all the scores in the internalisation questions in the moral identity scale) as well as the interaction of the internalisation score and the treatment variable. Hypothesis 3.2 corresponds to a significant positive interaction. Table 3, second column, shows that while the coefficient has the predicted sign, it is not significantly different from zero (coeff = 0.0228, p = 0.132).

|  | Symbolisation | Internalisation |
| --- | --- | --- |
| **Treatment Main Effect** | 1.027** | 0.0159 |
|  | (0.487) | (1.142) |
| **Moral identity main effect** | 0.192*** | -0.142 |
|  | (0.0655) | (0.0923) |
| **Interaction** | -0.163* | 0.0228 |
|  | (0.0887) | (0.132) |
| **Constant** | -1.319*** | 0.845** |
|  | (0.358) | (0.800) |
| **Observations** | 426 | 426 |

*p-Value<0.1, **p-value<.0.05, and ***p-value<.0.01.
(Standard errors in parentheses)

*Table 3. Logistic regressions with choosing the equal option as the dependent variable. Treatment is a dummy that takes value of 0 for baseline, moral identity is measured using the moral identity scale.*

**Discussion**

These results weakly support part of our hypothesis that preferences for either the injunctive norm or personal norm are linked to participants' moral identity. The regression with symbolisation scores weakly supports this and indicates that placing the injunctive norm on the efficient option makes those with higher symbolisation scores weakly less likely to choose the equal option. However, contrary to our predictions, we did not find clear evidence that people with higher internalisation scores are more likely to choose the equal choice, when this is framed as the right thing to do.

### General discussion

Here we reported three pre-registered studies (total N=1,799) testing the effect of nudging the personal and the injunctive norm in decisions that involve a trade-off between objective equality and efficiency. The first two studies provided evidence of a number of results: (i) both nudging the personal norm and the injunctive norm has an effect, and these effects are similar in size; (ii) when the personal norm and the injunctive norm are both nudged towards the same direction, then there is no additive effect; (iii) when the personal norm and the injunctive norm are nudged towards opposite directions, then some people tend to follow the personal norm and other people tend to follow the injunctive norm. In our third study, we tested whether these two classes of people, those who tend to follow the injunctive norm and those who tend to follow the personal norm, map onto the two subdimensions of Aquino and Reed's moral identity scale. We found only partial evidence of this: people higher in the symbolisation dimension are weakly more likely to follow the injunctive norm; however, we did not find any evidence that people higher in the internalisation dimension are more likely to follow the personal norm.

These results contribute to the literature along several dimensions. First of all, they provide a new example of a one-shot anonymous game in which nudging the injunctive norm has a significant effect. This is already a non-trivial result, since previous research has failed to find consistent and coherent evidence of a significant effect of nudging the injunctive norm on one-shot anonymous games: Bicchieri and Chavez (2010) found that manipulating the injunctive norm in the ultimatum game does have an effect, however Schram and Charness (2015) failed to show that manipulating the injunctive norm has an effect on the dictator game, and Dimant et al. (2020) failed to successfully manipulate the injunctive norm in the mind game. This also opens a general question about which are the one-shot anonymous games where the injunctive norm plays a role and those where it does not, and why. We believe this to be an important question for future research.

The second contribution of our results concerns the comparison between different types of norms. As detailed in the Introduction, there is very little work comparing the effect of norm-based interventions on one-shot economic games and, to the best of our knowledge, no studies have compared the personal and the injunctive norm. We did so in the context of the trade-off game. We found that, in this context, nudging each of these norms has an effect, and the effects are similar in size.

The third contribution of our research is the investigation of what happens when both interventions are used together. This is an important question from both the theoretical and the practical viewpoints, which has received little attention in the past. Moreover, the two papers we are aware of, focus on cases in which the norms are placed on different options (Bicchieri and Xiao, 2009; Capraro & Rand, 2018) rather than the same option. Consequently, very little is known about the additive properties of norm-based interventions. Our findings show that, in the context of a one-shot anonymous trade-off game, placing the injunctive norm and the personal norm on the same option does not have any additive effect. Yet, when they are placed on different options, some people tend to follow the personal norm, while others tend to follow the injunctive norm. One possible way to explain this asymmetry is by assuming that some people have general moral preferences and, therefore, when either the injunctive norm or the personal norm are placed on one option, then all these people move towards this option. Therefore, if both norms are placed on the same option, nothing else happens because all people with general moral preferences have already been moved towards this option. However, when one norm is placed on one option while the other norm is placed on the other option, then people turn to norm-specific preferences: those with a higher preference for the injunctive norm follows the injunctive norm, whereas those with a higher preference for the personal norm follows the personal norm. Future work should explore this hypothesis more in depth.

Our work also suggests another research question. What characteristics make people more likely to follow the injunctive or the personal norm? In Study 3 we tested the role of the two subdimensions of moral identity, symbolisation and internalisation. We found some evidence that people with higher symbolisation scores are more likely to follow the injunctive norm (the effect was marginally significant), but we did not find any clear evidence that people with higher internalisation scores are more likely to follow the personal norm. Future work could explore further individual correlates to personal and injunctive norm-following.

This work does have some limitations. The first limitation is that the stakes of our trade-off game were relatively small. Previous work suggests that stakes have little effect on people's behaviour in a number of economic games involving prosocial behaviours, at least when stakes are not as high as the order of magnitude of a month of salary (Forsythe et al., 1994; Carpenter, Verhoogen & Burks, 2005; Johansson, Stenman, Mahmud & Martinsson, 2005; Brañas-Garza et al., 2018; Larney, Rotella & Barclay, 2019). Therefore, we do believe this to be a minor limitation. The second limitation is that, in our experiments, the personal norm manipulation always targets the equal option, and never the efficient option. We made this experimental choice because, at baseline, people are slightly more likely to choose the efficient option compared to the equal option, therefore nudging the equal option means that there is more room for statistical changes, compared to nudging the efficient option. We believe this to be a minor issue as well, because previous work found that nudging the equal or the efficient option through a personal norm manipulation has similar effects (Capraro & Rand, 2018). A third limitation regards the generalisability of our results. Our experiments were conducted using a convenient WEIRD (Western Eductated Industrialized Rich Democratic) population of subjects based in the US (Heinrich, Heine, & Norenzayan, 2010). To what extent these results extend to other, possibly non-WEIRD populations, is an open question.

In sum, our results show that nudging the injunctive and the personal norm has an effect on the trade-off game; their effect is not additive; however, when the two nudges are misaligned, some people tend to follow the personal norm while others tend to follow the injunctive norm.

# Appendix: Experimental instructions

## Study 1

**Baseline condition ("BL")**

You are player A. You are playing a game with two other players. Player B and Player C. You get to make a choice (Player B and Player C do not make any decisions).

If you choose Option 1, you earn 13c, Player B earns 13c and player C earns 13c.

If you choose Option 2, you earn 13c, Player B earns 23c and player C earns 13c.

This is the only interaction you have with Player B and Player C. They will not have the opportunity to influence your gain in later parts of the HIT.

Now we are going to ask you some comprehension questions to make sure you have understood the decision problem. You will need to get them all correct to continue the survey. If you do not answer the comprehension questions correctly the survey will automatically end without providing a completion code. If that is the case please submit your HIT on mechanical Turk using your Turk ID instead of the completion code so that we can pay you the participation fee.

What choice should you make if you want all players to get the same payoff?

○ Option 1 (You earn 13 cents, Player B earns 13 cents, Player C earns 13 cents)

○ Option 2 (You earn 13 cents, Player B earns 23 cents, Player C earns 13 cents)

What choice should you make if you want to maximize the sum of the payoffs of all players?

○ Option 1 (You earn 13 cents, Player B earns 13 cents, Player C earns 13 cents)

○ Option 2 (You earn 13 cents, Player B earns 23 cents, Player C earns 13 cents)

(A skip logic eliminated all subjects who failed one or both comprehension questions. These participants were asked to submit their HIT using their TurkID at the place of the completion code, so that we could pay them the participation fee, but not the additional bonus.)

You have passed the comprehension questions. It is now time to make your decision.

○ Option 1 (You earn 13 cents, Player B earns 13 cents, Player C earns 13 cents)

○ Option 2 (You earn 13 cents, Player B earns 23 cents, Player C earns 13 cents)

**Personal norm equal condition ("PN_equal")**

You are player A. You are playing a game with two other players. Player B and Player C. You get to make a choice (Player B and Player C do not make any decisions).

If you choose to be fair, you earn 13c, Player B earns 13c and player C earns 13c.

If you choose to be unfair, you earn 13c, Player B earns 23c and player C earns 13c.

This is the only interaction you have with Player B and Player C. They will not have the opportunity to influence your gain in later parts of the HIT.

Now we are going to ask you some comprehension questions to make sure you have understood the decision problem. You will need to get them all correct to continue the survey. If you do not answer the comprehension questions correctly the survey will automatically end without providing a completion code. If that is the case please submit your HIT on mechanical Turk using your Turk ID instead of the completion code so that we can pay you the participation fee.

What choice should you make if you want all players to get the same payoff?

○ Be fair (You earn 13 cents, Player B earns 13 cents, Player C earns 13 cents)

○ Be unfair (You earn 13 cents, Player B earns 23 cents, Player C earns 13 cents)

What choice should you make if you want to maximize the sum of the payoffs of all players?

○ Be fair (You earn 13 cents, Player B earns 13 cents, Player C earns 13 cents)

○ Be unfair (You earn 13 cents, Player B earns 23 cents, Player C earns 13 cents)

(A skip logic eliminated all subjects who failed one or both comprehension questions. These participants were asked to submit their HIT using their TurkID at the place of the completion code, so that we could pay them the participation fee, but not the additional bonus.)

You have passed the comprehension questions. It is now time to make your decision.

○ Be fair (You earn 13 cents, Player B earns 13 cents, Player C earns 13 cents)

○ Be unfair (You earn 13 cents, Player B earns 23 cents, Player C earns 13 cents)

**Personal norm equal + injunctive norm equal condition ("PN_equal + IN_equal")**

You are player A. You are playing a game with two other players. Player B and Player C. You get to make a choice (Player B and Player C do not make any decisions).

If you choose to be fair, you earn 13c, Player B earns 13c and player C earns 13c.

If you choose to be unfair, you earn 13c, Player B earns 23c and player C earns 13c.

This is the only interaction you have with Player B and Player C. They will not have the opportunity to influence your gain in later parts of the HIT.

Now we are going to ask you some comprehension questions to make sure you have understood the decision problem. You will need to get them all correct to continue the survey. If you do not answer the comprehension questions correctly the survey will automatically end without providing a completion code. If that is the case please submit your HIT on mechanical Turk using your Turk ID instead of the completion code so that we can pay you the participation fee.

What choice should you make if you want all players to get the same payoff?

○ Be fair (You earn 13 cents, Player B earns 13 cents, Player C earns 13 cents)

○ Be unfair (You earn 13 cents, Player B earns 23 cents, Player C earns 13 cents)

What choice should you make if you want to maximize the sum of the payoffs of all players?

○ Be fair (You earn 13 cents, Player B earns 13 cents, Player C earns 13 cents)

○ Be unfair (You earn 13 cents, Player B earns 23 cents, Player C earns 13 cents)

(A skip logic eliminated all subjects who failed one or both comprehension questions. These participants were asked to submit their HIT using their TurkID at the place of the completion code, so that we could pay them the participation fee, but not the additional bonus.)

You have passed the comprehension questions. It is now time to make your decision.

Before making your decision, we can inform you that we asked 10 participants who played before you what they think the morally right thing to do is. Nine of these ten participants (90%) declared that choosing for all players to get the same payoffs is the morally right thing to do.

○ Be fair (You earn 13 cents, Player B earns 13 cents, Player C earns 13 cents)

○ Be unfair (You earn 13 cents, Player B earns 23 cents, Player C earns 13 cents)

**Personal norm equal + injunctive norm efficient condition ("PN_equal + IN_efficient")**

You are player A. You are playing a game with two other players. Player B and Player C. You get to make a choice (Player B and Player C do not make any decisions).

If you choose to be fair, you earn 13c, Player B earns 13c and player C earns 13c.

If you choose to be unfair, you earn 13c, Player B earns 23c and player C earns 13c.

This is the only interaction you have with Player B and Player C. They will not have the opportunity to influence your gain in later parts of the HIT.

Now we are going to ask you some comprehension questions to make sure you have understood the decision problem. You will need to get them all correct to continue the survey. If you do not answer the comprehension questions correctly the survey will automatically end without providing a completion code. If that is the case please submit your HIT on mechanical Turk using your Turk ID instead of the completion code so that we can pay you the participation fee.

What choice should you make if you want all players to get the same payoff?

- ○ Be fair (You earn 13 cents, Player B earns 13 cents, Player C earns 13 cents)

- ○ Be unfair (You earn 13 cents, Player B earns 23 cents, Player C earns 13 cents)

What choice should you make if you want to maximize the sum of the payoffs of all players?

- ○ Be fair (You earn 13 cents, Player B earns 13 cents, Player C earns 13 cents)

- ○ Be unfair (You earn 13 cents, Player B earns 23 cents, Player C earns 13 cents)

(A skip logic eliminated all subjects who failed one or both comprehension questions. These participants were asked to submit their HIT using their TurkID at the place of the completion code, so that we could pay them the participation fee, but not the additional bonus.)

You have passed the comprehension questions. It is now time to make your decision.

Before making your decision, we can inform you that we asked 10 participants who played before you what they think the morally right thing to do is. Nine of these ten participants (90%) declared that choosing to maximize the sum of the payoffs of all players is the morally right thing to do.

- ○ Be fair (You earn 13 cents, Player B earns 13 cents, Player C earns 13 cents)

- ○ Be unfair (You earn 13 cents, Player B earns 23 cents, Player C earns 13 cents)

**Injunctive norm efficient condition ("IN_efficient")**

You are player A. You are playing a game with two other players. Player B and Player C. You get to make a choice (Player B and Player C do not make any decisions).

If you choose Option 1, you earn 13c, Player B earns 13c and player C earns 13c.

If you choose Option 2, you earn 13c, Player B earns 23c and player C earns 13c.

This is the only interaction you have with Player B and Player C. They will not have the opportunity to influence your gain in later parts of the HIT.

Now we are going to ask you some comprehension questions to make sure you have understood the decision problem. You will need to get them all correct to continue the survey. If you do not answer the comprehension questions correctly the survey will automatically end without providing a completion code. If that is the case please submit your HIT on mechanical Turk using your Turk ID instead of the completion code so that we can pay you the participation fee.
What choice should you make if you want all players to get the same payoff?

- ○ Option 1 (You earn 13 cents, Player B earns 13 cents, Player C earns 13 cents)

- ○ Option 2 (You earn 13 cents, Player B earns 23 cents, Player C earns 13 cents)

What choice should you make if you want to maximize the sum of the payoffs of all players?

- ○ Option 1 (You earn 13 cents, Player B earns 13 cents, Player C earns 13 cents)

- ○ Option 2 (You earn 13 cents, Player B earns 23 cents, Player C earns 13 cents)

(A skip logic eliminated all subjects who failed one or both comprehension questions. These participants were asked to submit their HIT using their TurkID at the place of the completion code, so that we could pay them the participation fee, but not the additional bonus.)

You have passed the comprehension questions. It is now time to make your decision.

Before making your decision, we can inform you that we asked 10 participants who played before you what they think the morally right thing to do is. Nine of these ten participants (90%) declared that choosing to maximize the sum of the payoffs of all players is the morally right thing to do.

- ○ Option 1 (You earn 13 cents, Player B earns 13 cents, Player C earns 13 cents)

- ○ Option 2 (You earn 13 cents, Player B earns 23 cents, Player C earns 13 cents)

## Study 2

**Baseline condition ("BL")**

You are player A. You are playing a game with two other players. Player B and Player C. You get to make a choice (Player B and Player C do not make any decisions).

If you choose Option 1, you earn 5c, Player B earns 5c and player C earns 5c.

If you choose Option 2, you earn 5c, Player B earns 10c and player C earns 5c.

This is the only interaction you have with Player B and Player C. They will not have the opportunity to influence your gain in later parts of the HIT.

Now we are going to ask you some comprehension questions to make sure you have understood the decision problem. You will need to get them all correct to continue the survey. If you do not answer the comprehension questions correctly the survey will automatically end without providing a completion code. If that is the case please submit your HIT on mechanical Turk using your Turk ID instead of the completion code so that we can pay you the participation fee.

What choice should you make if you want all players to get the same payoff?

○ Option 1 (You earn 5cents, Player B earns 5 cents, Player C earns 5 cents)

○ Option 2 (You earn 5 cents, Player B earns 10 cents, Player C earns 5 cents)

What choice should you make if you want to maximize the sum of the payoffs of all players?

○ Option 1 (You earn 5 cents, Player B earns 5 cents, Player C earns 5 cents)

○ Option 2 (You earn 5 cents, Player B earns 10 cents, Player C earns 5 cents)

(A skip logic eliminated all subjects who failed one or both comprehension questions. These participants were asked to submit their HIT using their TurkID at the place of the completion code, so that we could pay them the participation fee, but not the additional bonus.)

You have passed the comprehension questions. It is now time to make your decision.

○ Option 1 (You earn 5 cents, Player B earns 5 cents, Player C earns 5 cents)

○ Option 2 (You earn 5 cents, Player B earns 10 cents, Player C earns 5 cents)

**Personal norm equal condition ("PN_equal")**

You are player A. You are playing a game with two other players. Player B and Player C. You get to make a choice (Player B and Player C do not make any decisions).

If you choose to be fair, you earn 5c, Player B earns 5c and player C earns 5c.

If you choose to be unfair, you earn 5c, Player B earns 10c and player C earns 5c.

This is the only interaction you have with Player B and Player C. They will not have the opportunity to influence your gain in later parts of the HIT.

Now we are going to ask you some comprehension questions to make sure you have understood the decision problem. You will need to get them all correct to continue the survey. If you do not answer the comprehension questions correctly the survey will automatically end without providing a completion code. If that is the case please submit your HIT on mechanical Turk using your Turk ID instead of the completion code so that we can pay you the participation fee.

What choice should you make if you want all players to get the same payoff?

- ○ Be fair (You earn 5 cents, Player B earns 5 cents, Player C earns 5 cents)

- ○ Be unfair (You earn 5 cents, Player B earns 10 cents, Player C earns 5 cents)

What choice should you make if you want to maximize the sum of the payoffs of all players?

- ○ Be fair (You earn 5 cents, Player B earns 5 cents, Player C earns 5 cents)

- ○ Be unfair (You earn 5 cents, Player B earns 10 cents, Player C earns 5 cents)

(A skip logic eliminated all subjects who failed one or both comprehension questions. These participants were asked to submit their HIT using their TurkID at the place of the completion code, so that we could pay them the participation fee, but not the additional bonus.)

You have passed the comprehension questions. It is now time to make your decision.

- ○ Be fair (You earn 5 cents, Player B earns 5 cents, Player C earns 5 cents)

- ○ Be unfair (You earn 5 cents, Player B earns 10 cents, Player C earns 5 cents)

**Injunctive norm equal condition ("IN_equal")**

You are player A. You are playing a game with two other players. Player B and Player C. You get to make a choice (Player B and Player C do not make any decisions).

If you choose Option 1, you earn 5c, Player B earns 5c and player C earns 5c.

If you choose Option 2, you earn 5c, Player B earns 10c and player C earns 5c.

This is the only interaction you have with Player B and Player C. They will not have the opportunity to influence your gain in later parts of the HIT.

Now we are going to ask you some comprehension questions to make sure you have understood the decision problem. You will need to get them all correct to continue the survey. If you do not answer the comprehension questions correctly the survey will automatically end without providing a completion code. If that is the case please submit your HIT on mechanical Turk using your Turk ID instead of the completion code so that we can pay you the participation fee.

What choice should you make if you want all players to get the same payoff?

- ○ Option 1 (You earn 5 cents, Player B earns 5 cents, Player C earns 5 cents)

- ○ Option 2 (You earn 5 cents, Player B earns 10 cents, Player C earns 5 cents)

What choice should you make if you want to maximize the sum of the payoffs of all players?

- ○ Option 1 (You earn 5 cents, Player B earns 5 cents, Player C earns 5 cents)

- ○ Option 2 (You earn 5 cents, Player B earns 10 cents, Player C earns 5 cents)

(A skip logic eliminated all subjects who failed one or both comprehension questions. These participants were asked to submit their HIT using their TurkID at the place of the completion code, so that we could pay them the participation fee, but not the additional bonus.)

You have passed the comprehension questions. It is now time to make your decision.

Before making your decision, we can inform you that we asked 10 participants who played before you what they think the morally right thing to do is. Nine of these ten participants (90%) declared that choosing for all players to get the same payoffs is the morally right thing to do.

- ○ Option 1 (You earn 5 cents, Player B earns 5 cents, Player C earns 5 cents)

- ○ Option 2 (You earn 5 cents, Player B earns 10 cents, Player C earns 5 cents)

**Study 3**

**Baseline condition ("BL")**

You are player A. You are playing a game with two other players. Player B and Player C. You get to make a choice (Player B and Player C do not make any decisions).

If you choose Option 1, you earn 5c, Player B earns 5c and player C earns 5c.

If you choose Option 2, you earn 5c, Player B earns 10c and player C earns 5c.

This is the only interaction you have with Player B and Player C. They will not have the opportunity to influence your gain in later parts of the HIT.

Now we are going to ask you some comprehension questions to make sure you have understood the decision problem. You will need to get them all correct to continue the survey. If you do not answer the comprehension questions correctly the survey will automatically end without providing a completion code. If that is the case please submit your HIT on mechanical Turk using your Turk ID instead of the completion code so that we can pay you the participation fee.

What choice should you make if you want all players to get the same payoff?

- ○ Option 1 (You earn 5cents, Player B earns 5 cents, Player C earns 5 cents)

- ○ Option 2 (You earn 5 cents, Player B earns 10 cents, Player C earns 5 cents)

What choice should you make if you want to maximize the sum of the payoffs of all players?

- ○ Option 1 (You earn 5 cents, Player B earns 5 cents, Player C earns 5 cents)
- ○ Option 2 (You earn 5 cents, Player B earns 10 cents, Player C earns 5 cents)

(A skip logic eliminated all subjects who failed one or both comprehension questions. These participants were asked to submit their HIT using their TurkID at the place of the completion code, so that we could pay them the participation fee, but not the additional bonus.)

You have passed the comprehension questions. It is now time to make your decision.

- ○ Option 1 (You earn 5 cents, Player B earns 5 cents, Player C earns 5 cents)
- ○ Option 2 (You earn 5 cents, Player B earns 10 cents, Player C earns 5 cents)

**Personal norm equal + injunctive norm efficient condition ("PN_equal + IN_efficient")**

You are player A. You are playing a game with two other players. Player B and Player C. You get to make a choice (Player B and Player C do not make any decisions).

If you choose to be fair, you earn 5c, Player B earns 5c and player C earns 5c.

If you choose to be unfair, you earn 5c, Player B earns 10c and player C earns 5c.

This is the only interaction you have with Player B and Player C. They will not have the opportunity to influence your gain in later parts of the HIT.

Now we are going to ask you some comprehension questions to make sure you have understood the decision problem. You will need to get them all correct to continue the survey. If you do not answer the comprehension questions correctly the survey will automatically end without providing a completion code. If that is the case please submit your HIT on mechanical Turk using your Turk ID instead of the completion code so that we can pay you the participation fee.

What choice should you make if you want all players to get the same payoff?

- ○ Be fair (You earn 5 cents, Player B earns 5 cents, Player C earns 5 cents)
- ○ Be unfair (You earn 5 cents, Player B earns 10 cents, Player C earns 5 cents)

What choice should you make if you want to maximize the sum of the payoffs of all players?

- ○ Be fair (You earn 5 cents, Player B earns 5 cents, Player C earns 5 cents)
- ○ Be unfair (You earn 5 cents, Player B earns 10 cents, Player C earns 5 cents)

(A skip logic eliminated all subjects who failed one or both comprehension questions. These participants were asked to submit their HIT using their TurkID at the place of the completion code, so that we could pay them the participation fee, but not the additional bonus.)

You have passed the comprehension questions. It is now time to make your decision.

Before making your decision, we can inform you that we asked 10 participants who played before you what they think the morally right thing to do is. Nine of these ten participants (90%) declared that choosing to maximize the sum of the payoffs of all players is the morally right thing to do.

○ Be fair (You earn 5 cents, Player B earns 5 cents, Player C earns 5 cents)

○ Be unfair (You earn 5 cents, Player B earns 10 cents, Player C earns 5 cents)

**Moral Identity (common to all Study 3 participants who correctly answered the comprehension questions)**

Listed below are some characteristics that might describe a person: caring, compassionate, fair, friendly, generous, helpful, hardworking, honest, kind. The person with these characteristics could be you or it could be someone else. For a moment, visualize in your mind the kind of person who has these characteristics. Imagine how that person would think, feel, and act. When you have a clear image of what this person would be like, answer the following questions.

|        | Strongly disagree | Neither agree nor disagree | Strongly agree |
|--------|-------------------|----------------------------|----------------|
|        | 0  1  2  3  4  5  6  7  8  9  10 | | |

| | |
|---|---|
| It would make me feel good to be a person who has these characteristics. () | 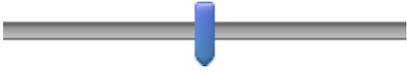 |
| Being someone who has these characteristics is an important part of who I am. () | 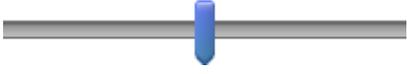 |
| I often wear clothes that identify me as having these characteristics. () | 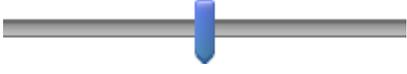 |
| I would be ashamed to be a person who had these characteristics. () | 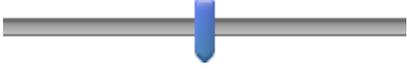 |
| The types of things I do in my spare time (e.g., hobbies) clearly identify me as having these characteristics () | 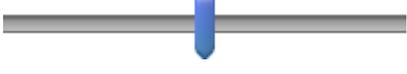 |
| The kinds of books and magazines that I read identify me as having these characteristics. () | 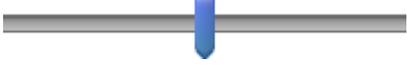 |
| Having these characteristics is not really important to me. () | 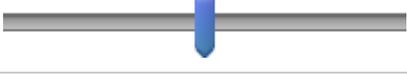 |
| The fact that I have these characteristics is communicated to others by my membership in certain organizations. () | 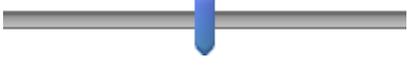 |
| I am actively involved in activities that communicate to others that I have these characteristics. () | 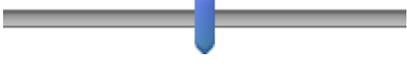 |
| I strongly desire to have these characteristics. () | 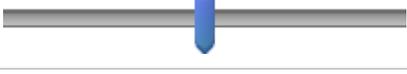 |